# Collimators

*Slawomir Wronka*
National Centre for Nuclear Research, Otwock, Poland

**Abstract**
The collimator system of a particle accelerator must efficiently remove stray particles and provide protection against uncontrolled losses. In this article, the basic design concepts of collimators and some realizations are presented.

## 1  Introduction

In all types of linear and circular accelerators, collimators are required to narrow the beam of particles. Owing to differences in the construction of the various types of accelerators, there are various approaches to beam collimation. In linacs, it is important to collimate the beam before the target or before or within the transfer line. In this type of machine, the beam interacts with the collimating system only once. In contrast, in synchrotrons and accumulator rings, the collimating system affects the beam parameters continuously and the proper selection of collimator locations is a more complicated problem.

Historically, collimators have been used in *hadron machines* to reduce the radiation background at the experimental sites. However, new machines, owing to the high energy and high luminosity of the beam and also the use of superconducting technologies, require sophisticated collimation systems for beam cleaning and machine protection. In modern accelerators, high intensities occur not only in the beam core but also in the beam halo created by particle migration due to collisions, beam–gas interactions, or nonlinearities in the magnetic fields. Lost particles originating from beam tails create uncontrolled losses, emittance growth, activation of accelerator components, heat deposition, and potential quenches in superconducting magnets. Properly designed collimator sections allow controlled, cumulative deposition of the losses in well-known, prepared locations, and thus minimize the impact of radiation on equipment.

In general, and regardless of the type of machine (synchrotron or linac), the objectives of collimators can be defined as follows:

– to obtain low uncontrolled beam loss;
– to minimize the halo around the proton or ion beam;
– to minimize the activation of downstream beam line components;
– to allow faster access to the machine and to experimental sites;
– to protect the machine itself against damage.

## 2  Types of collimators

The typical components of the collimation system can be categorized by their function, as follows.

### 2.1  Jaws

These typically consist of two solid blocks and are used for efficient beam cleaning. The thickness is high enough to stop impacting particles, although the production of secondary particles is possible (see Fig. 1). The correct selection of the material is important to avoid damage in the case of high power deposition. For ultrasmall beams (as in the LHC; see Fig. 2), high precision and stringent tolerances are required. Primary jaws are often followed by secondary and tertiary collimators.

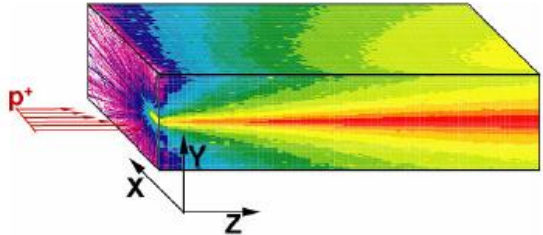

**Fig. 1:** Schematic drawing of a collimator jaw [1]

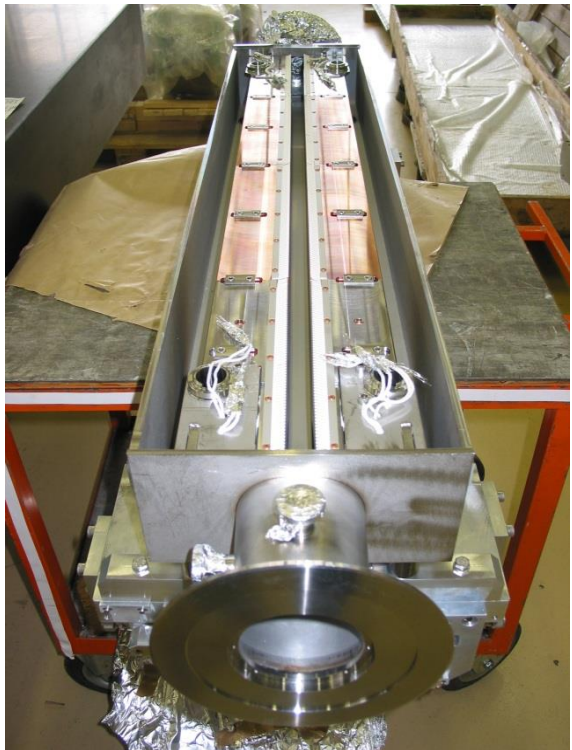

**Fig. 2:** Jaws of the LHC accelerator [2]

A heavy charged particle traversing matter loses energy primarily through the ionization and excitation of atoms [3]. The energy transferred may be sufficient to knock an electron out of an atom and thus ionize it, or it may leave the atom in an excited, non-ionized state. Because the mass of a heavy charged particle is thousands of times larger than the electron mass, the particle can transfer only a small fraction of its energy in a single electronic collision. The deflection of the particle in the collision is negligible. Thus, a heavy charged particle travels in an almost straight path through matter, losing energy almost continuously in small amounts through collisions with atomic electrons, leaving ionized and excited atoms in its wake. The average linear rate of energy loss of a heavy charged particle in a medium (expressed, for example, in MeV·cm$^{-1}$) is of fundamental importance in radiation physics and dosimetry. This quantity, denoted by $-dE/dx$, is called the stopping power of the medium for the particle.

The range of a charged particle is the distance it travels before coming to rest. The reciprocal of the stopping power gives the distance travelled per unit energy loss. Therefore, the range $R(T)$ of a particle of kinetic energy $T$ is the integral of this quantity down to zero energy:

$$R(T) = \int_0^T \left(-\frac{dE}{dx}\right)^{-1} dE. \qquad (1)$$

In practice, Monte Carlo codes are used to calculate the minimum length required for a set of jaws, for example MCNP [4], Fluka [5], and Geant4 [6].

### 2.2 Scrapers

Thin objects (e.g. foils) are typically used for beam shaping and diagnostics. Scrapers can be used together with magnets located in the beam line. For example, in the case of H⁻ particles, scrapers change the charge and the magnets then naturally bend the particles out of the beam line [7]. An example of a scraper is presented in Fig. 3.

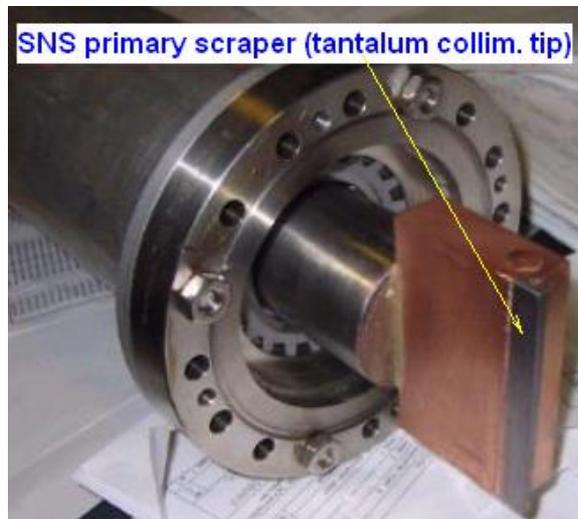

**Fig. 3:** Scraper at the Spallation Neutron Source (SNS) [1]

### 2.3 Absorbers

Movable absorbers can be quite similar in design to jaws. The main goal of these elements is to absorb miskicked beam or products of particle-induced showers. In comparison with jaws, larger gaps and relaxed tolerances are used.

### 2.4 Additional equipment for precise set-up and alignment

Beam loss monitors (BLMs) are used for precise alignment of collimators. When a jaw or scraper 'touches' the beam, a stronger secondary shower is produced, which is detected by the BLMs. Such a procedure can be repeated for each element and each side of the beam.

## 3 Practical realization of collimator systems

Typically, the whole of a collimation system must be carefully calculated and designed. It is important to define the length of the collimating tube; the shape, location, and number of collimators (along the entire lattice and in each set); and the aperture size of the primary, secondary, and tertiary collimators. The material of the collimators must effectively stop the particles, survive the resulting heat load, and receive as low an activation as possible. Systematic material studies must be performed to verify the ability of materials to withstand thermomechanical shock and to measure other parameters, such as thermal expansion coefficients and radiation resistance for long exposures. Finally, the requirements for a stationary or movable collimator imposed by the design of the accelerator or detector must be taken into account; for example, one must analyse if any movement is required and, if so, what algorithm will be used to control the movement.

In such calculations and engineering design work, one has to take the following into account:

– the beam power;

– all potential losses and beam halo;

– showers in the collimators and other equipment, and electron clouds;

– material behaviour and beam-induced damage, and elastic and inelastic deformation;

– the possible use of coatings on the base material to modify its parameters;

– the need for precise mechanical movement and highly efficient cooling;

– the radioactivity levels in the collimator regions (which affect both the materials and personnel);

– tolerance requirements;

– a risk analysis of the potential failure scenarios (for example, the particle trajectories may change as a result of a klystron misfire or a magnet failure).

Such projects are almost always very complicated and require tight co-operation among the people responsible for different accelerator and detector sections. For example, the activation of the collimators themselves and of downstream elements, together with the shielding requirements for each collimation section, must be taken into account (see [8, 9] for a description of an example). It is necessary to study a 'distributed collimation' approach, where small collimators are located in many places sandwiched between other elements, and compare it with a 'bulk collimation' philosophy, in which beam collimation is done at only two or three locations. The output of such complex studies should define the collimator geometry, the collimation material, and the cooling requirements for the various levels of intercepted power that will be encountered.

The future development of collimators will take account of the possibility of using new materials to achieve longer lifetimes in an ultrahigh-radiation environment and under high thermal and mechanical stresses. Some new concepts are based on multiuse collimators, an example of which is presented in Fig. 4.

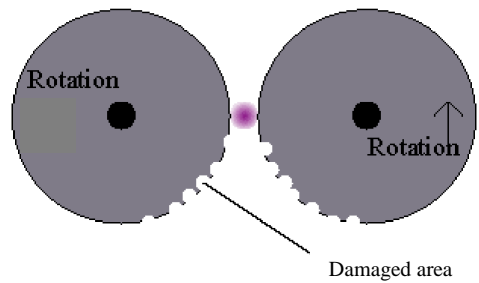 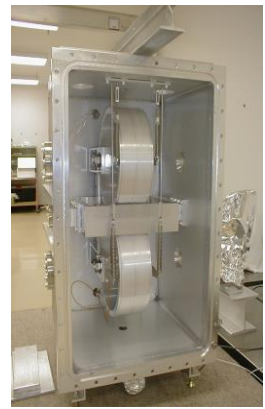

**Fig. 4:** Concept of a multiuse rotating-wheel collimator. This is an example of a 'consumable spoiler' to be used at the SLAC NLC [9, 10].

## 4 Summary

A general introduction to collimators, which are very important elements of accelerators, has been presented in this paper. A short description of different types of collimators and their roles was given, with examples from the LHC and SNS. In modern high-power hadron machines, collimators are of principal importance for ensuring the safety of the machine itself and of the detectors downstream of the beam, as well as being an important part of the radiation safety system.